# Electromelting of Confined Monolayer Ice

Hu Qiu, Wanlin Guo[†]

State Key Laboratory of Mechanics and Control of Mechanical Structures and the Key Laboratory for Intelligent Nano Materials and Devices of MoE,

Institute of Nanoscience, Nanjing University of Aeronautics and Astronautics,

29 Yudao Street, Nanjing 210016, China

**Abstract**

In sharp contrast to the prevailing view that electric fields promote water freezing, here we show by molecular dynamics simulations that monolayer ice confined between two parallel plates can melt into liquid water under perpendicularly applied electric field. The melting temperature of the monolayer ice decreases with the increasing strength of the external field due to field-induced disruption of the water-wall interaction induced well-ordered network of hydrogen bond. This electromelting process should add an important new ingredient to the physics of water.

PACS numbers: 61.20.Ja, 64.70.D-, 68.18.Fg

[†] Correspondence should be addressed to W. G. (wlguo@nuaa.edu.cn)

Phase transition between liquid and solid water is of fundamental importance in environmental, biological and industrial technologies. Although the normal freezing point of bulk water is 0 °C, extremely pure liquid water cannot freeze even being cooled down to -40 °C [1]. The reason is that the high activation energy for homogeneous nucleation impedes the spontaneous formation of stable seed nucleus for crystallization. It has been widely reported that the crystallization of liquid bulk water into cubic ice can be promoted by electric fields, so-called electrofreezing [2-6]. Freezing to a crystalline phase can also occur spontaneously when water is subjected to an extremely confined state due to the water-wall interaction. A number of computational [7-10] and experimental [11] studies have demonstrated the spontaneous freezing of few-layer ice confined between two parallel plates separated by nanometer gaps. Thus it is thermodynamically more favorable to freeze liquid water confined within a nanogap [12-14] or absorbed at charged surfaces [15] by external electric fields due to the confinement-reduced translational entropy of water. All these theoretical simulations and experimental facts form a prevailing view that electric fields always promote the freezing of both the bulk and confined water through electric-field-induced realignment of the dipoles of water molecules. However, how an applied electric field plays its role in the reverse process to freezing (i.e., ice melting) has never been reported.

In this letter, we show by molecular dynamics (MD) simulations that the confined monolayer ice between two plates can melt into liquid water under a perpendicularly applied electric field. The melting temperature of the monolayer ice can be significantly reduced by stronger electric fields. This unexpected electric-field-induced melting of ice, which can be named as electromelting, is mainly due to the interruption of hydrogen bond (H-bond) network between water molecules under the perpendicular electric field.The simulation system consists of 778 water molecules confined between two parallel plates with area of ~7.4×7.4 nm$^2$ along the *x-y* plane with a spacing distance *d*, as measured between the centers of plate atoms (Fig. 1, bottom inset). The plates consist of a triangular arrangement of monolayer atoms, with the neighboring bond length being fixed at 0.23 nm. It has been shown that accounting for polarizability of water in MD simulations under electric fields can affect the position of the phase boundary, but does not change the existence of each water phase in the phase diagram [16], so the non-polarizable TIP5P model [17] will be used in our simulations without specific statement. Although it was proposed that molecules in bulk water can be instantaneously dissociated by electric fields ranging from 3.5 to 10 V/nm [18], our density functional theory calculations show that such dissociation cannot occur in our confined system under the perpendicularly applied electric fields [19]. The water-wall interactions were treated by a 6-12 Lennard-Jones (LJ) potential with the parameters: $\sigma_{(O\text{-wall})}$ = 0.316 nm, $\sigma_{(H\text{-wall})}$ = 0.284 nm and $\varepsilon_{(O\text{-wall})}$ = 0.831 kJ/mol, $\varepsilon_{(H\text{-wall})}$ = 0.415 kJ/mol. These parameters correspond approximately to the van der Waals (vdW) interaction between a water molecule and quartz (SiO$_2$) surface [7, 8] Periodic boundary conditions were applied in the *x* and *y* directions to model the confinement between two infinite plates, and a large vacuum layer along the *z* direction was used to separate adjacent images of the simulation system to avoid any unphysical interactions. The MD simulations were performed using a time step of 2 fs with the GROMACS 4.0.5 software package [20]. When the electric field is weaker than 5 V/nm, the *NP$_{xy}$T* ensemble was adopted, with the *Berendsen* algorithm [21] being used to control the temperature *T* at 300K and the lateral pressure $P_{xy}$ at 1 bar. For stronger electric fields, the *NVT* ensemble was used instead to prevent the water molecules from leaving the confined area via the gap induced by the possible mismatch between the fixed confining plates and the expansible periodic simulation box along the *x-y* plane. The lateral pressure for liquid water, $P_{xy}$, is found to be nearly constant in *NVT* simulations for $E < 26.5$ V/nm [19]. The nonbonded interactions were treated with a twin range cutoff of 0.9 and 1.4 nm. All the other simulation parameters were the same as used in Refs. [7] and [8]. Under a given electric field, the system was equilibrated for 2 ns, and then the system evolved for 8 ns for data analysis. As adopted in previous simulations [2-6, 13, 14], the field strength used in our MD simulations is from several V/nm to several tens of V/nm. This range of electric field is at least $1 \times 10^3$ stronger than the external field applied in electrofreezing experiments [12, 22], but comparable to those experienced by water molecules near the surfaces of certain types of biopolymers [23] or within cracks in amino acid crystals [24]. In our model, two water molecules were considered to form H-bond if their O…O distance is shorter than 3.5 Å and simultaneously the angle O–H…O is larger than 150°.

Previous MD simulations have shown that both monolayer ice and liquid water can form at room temperature between plates dependent on their separating distances *d* [7]. The solidification of nanoconfined water was also observed experimentally at room temperature [11, 25]. Consistent with the previous MD study [7]., our simulation results suggest that the lateral diffusion coefficient of the monolayer ice at $d = 0.79$ nm (blue curve) is about 3 orders lower in magnitude than the liquid water at $d = 0.81$ nm (red curve) in the absence of electric fields ($E = 0$ in Fig. 1). When a perpendicular electric field *E* is applied, the mobility of the monolayer ice remains nearly unchanged for $E < 2.5$ V/nm (see inset I in Fig. 1), with the lateral diffusion coefficient fluctuating slightly in the order of $\sim 10^{-8}$ cm$^2$/s. When the electric field enters the range of $2.5 \leq E \leq 3.75$ V/nm, the diffusion coefficient of the monolayer ice increases more and more quickly with increasing *E*. When the electric field approaches 3.8 V/nm, a sudden jump occurs in the lateral diffusion coefficient of the monolayer at spacing $d = 0.79$ nm from the order of $10^{-7}$ cm$^2$/s to the order of

~$10^{-5}$ cm$^2$/s, merging with that of the liquid water at spacing $d$ = 0.81 nm. This sudden jump in diffusion coefficient means that the monolayer ice melts into a liquid state (i.e., electromelting), as also shown by inset II in Fig. 1. In the range of 3.8 V/nm ≤ $E$ < 25 V/nm, the system at spacing $d$ = 0.79 nm represents a stable liquid configuration. In comparison, the water layer between plates of spacing $d$ = 0.81 nm remains in its liquid state in the whole range of 0 ≤ $E$ < 30 V/nm. Further increasing $E$ will lead to a sharp drop in the diffusion coefficient profiles for both plate separations, indicating a transition into a specific crystal-like structure, as shown by inset III in Fig. 1 at $E$ = 50 V/nm, which exhibits strongly reduced translational and rotational dynamics with enhanced structural order [26]. The critical field strength for this liquid-to-crystal transition slightly over 35 V/nm at spacing $d$ = 0.81 nm is some stronger than that at spacing $d$ = 0.79 nm, about 27 V/nm. It is interesting that similar electric-field-induced liquid-to-crystal transition can also occur in bulk water at a critical field strength of 30~40 V/nm [5, 6]. To check the reliability of the TIP5P water model, we further conducted a series of MD simulations with TIP4P water model [27]. The yielded diffusion coefficient profile in ref. [19] exhibits similar behavior of phase transition, even though the detailed value of the critical field strength shows some difference, indicating the robustness of the demonstrated trend of phase transitions.

The three above demonstrated water phases show distinguishing structural features as shown by Fig. 2. Figures 2(a) and 2(b) show the transverse distributions of the density of oxygen and hydrogen atoms, respectively, at $d$ = 0.79 nm under different electric fields. Without electric fields, water molecules in the ice monolayer confined between the two plates take a puckered structure with their oxygen atoms waving between two planes at $z$ ≈ ±0.0775 nm connected by hydrogen bonds, as shown by Fig. 2(a) and the upper panel of Fig. 2(c). This unique symmetric distribution of layered water about the middle plane at $z = 0$ is mainly attributed to the ordering effect caused by the water-wall interaction. When an external field of 5 V/nm is applied, both the planes are dragged downward approaching the lower confining plate, but the downward displacement of the lower oxygen plane (~0.0225 nm) is about 9 times of that of the upper oxygen plane (~0.0025 nm), as shown by the lower panel in Fig. 2(c). This electric-field-induced shift leads to a 12.9% increment in the spacing between the two oxygen planes. The detailed hydrogen distribution profiles at different fields are presented in Fig. 2(b). For the solid phase in the field-free state at $E = 0$, similar to the oxygen distribution, the hydrogen atoms are also located symmetrically with respect to the middle plane of $z = 0$. When an external field of 5 V/nm is applied, the liquid water completely loses the water-wall interaction induced symmetry in hydrogen distribution, as water molecules tend to point their hydrogen atoms upward in response to the electric field (see also inset II in Fig. 1).

We then further explored the structural properties of the monolayer ice and the electric-field-induced liquid phases by examining their oxygen-oxygen radial distribution functions (RDFs) $g_{OO}$ of water shown in Fig. 2(d). In the field-free case of $E = 0$, the well defined density maxima and minima validate the high structural order of the ice phase due to the water-wall interaction, in which each water molecule exhibits small amplitude fluctuation around its fixed position (see also inset I of Fig. 1). When an electric field of $E = 5$ V/nm is applied, the system loses its long-range structural order and changes into a liquid phase, as indicated by the decreased height of the second and third peaks of $g_{OO}$. The inherent structure of this liquid phase shows a disordered H-bond network (also see the inset II of Fig. 1). This electromelting of the confined monolayer ice can be further verified by the probability distribution of the angle $\varphi$ between water molecule dipole orientation ***M*** and the *x* axis, as shown in the inset of Fig. 2(d). In the field-free case, there is a clear dipole orientation preference with two peaks ($\varphi$ = 45° and 135°) for the monolayer ice due to the confinement of the walls. At $E$ = 5 V/nm, there is no evident peak in the $\varphi$ distribution, confirming the loss of structural order. It should be noted that significant discrepancy exists between the density profiles for the electric-field-induced liquid water at $d$ = 0.79 nm [red dotted lines in Figs. 2(a) and 2(b)] and that at $d$ = 0.81 V/nm in the field-free state [black solid lines in Figs. 2(a) and 2(b)], indicating that their inherent configurations are not the same. However, the nearly identical structural disorder of them, as indicated by the similar RDF and $\varphi$ distributions [red dotted lines and black solid lines in Fig. 2(d)], clearly shows their common liquid water nature.

The process of electric-field-induced symmetry breaking of the confined water between the two walls can be reflected more directly by monitoring the change in oxygen-wall distance with respect to electric field as shown in Fig. 3a, where the distance $D_1$ between the upper water plane and wall and $D_2$ between the lower water plane and wall, as well as their difference $D_1$-$D_2$ are presented as a function of the applied electric field. For a weak electric field with $E \leq 3.75$ V/nm, the symmetric structure of water remains nearly unchanged with $D_1 = D_2 = 0.3175$ nm and $D_1$-$D_2$ = 0. Although $D_1$ remains constant until $E$ = 4 V/nm, a sharp drop in $D_2$ occurs at $E$ = 3.8 V/nm, breaking the symmetry with significant increase in $D_1$-$D_2$. The evolution of the structural order of confined water with increasing field determined by the competition between the water-wall interaction and the applied electric field can be further explored by counting the average number of H-bonds formed by a water molecule with its neighboring water molecules in the same oxygen plane, $N_{intra}$, and the number of H-bonds formed by a water molecule with the water molecules in another oxygen plane, $N_{inter}$, as shown in Fig. 3b. It is found that both $N_{intra}$ and $N_{inter}$, as well as their sum $N_{total}$, are almost constant for $E < 3.75$ V/nm around 0.4, 1.7 and 2.1, respectively, consistent with the range of electric field for the ice phase as shown in Fig. 1. When $E$ is further increased to 3.8 V/nm, a large sudden drop in $N_{\text{inter}}$ and a smaller sharp jump in $N_{\text{intra}}$ occur, resulting in a small drop in $N_{\text{total}}$. The large drop in $N_{\text{inter}}$ suggests that the interplane H-bonds of the monolayer ice are largely interrupted by the strong electric field, leading to the transition to the liquid phase. It is also found that after the transition the two numbers ($N_{\text{intra}}$ and $N_{\text{inter}}$) in the liquid phase became closer to each other than in the monolayer ice [Fig. 3(b)], further confirming the loss of structural order of the system. Figure 3(c) shows the probability distribution $P(t)$ of the lifetime of H-bond under different electric fields. It is found that $P(t)$ in the liquid water at $E = 5$ V/nm decays significantly faster than that in the ice phase at $E = 0$. This further suggests that the electromelting is due to the electric-field-induced breaking of the symmetric distribution and structural order of confined water determined by the water-wall interaction. It was widely believed that the favorable energetic interactions of electric fields with the dipole moments of water molecules can reduce the entropy of liquid water due to the electric-field-induced restriction of the orientational degrees of freedom [13]. If the entropy reduction is large enough, water would transform to a crystalline phase (electrofreezing), as has been demonstrated in a computer simulation of confined liquid water under an external field applied in a direction parallel to the confining plates [13]. In contrast, under external fields with similar magnitude but perpendicular to the confining plates, what we demonstrate here is indeed an electric-field-induced melting (electromelting) process involving an increase in entropy. Thus the perpendicaular electric fields applied here do not help to align the dipoles of water molecules to stabilize the H-bond network, but rather, to break the structural order caused by the water-wall interaction and lead to the ice-to-liquid transition.

The dependence of the electromelting point, namely the temperature at which the melting occurs at a given electric field, should be an interesting and important issue to be investigated. Figure 4 presents our calculated temperature dependence of lateral diffusion coefficient at various field strengths. In the field-free case, a sharp jump in diffusion coefficient of the system at $T_m = 325$ K indicates that it melts into a liquid phase at this critical temperature, yielding the melting point. With increasing field strength, the electromelting point decreases gradually from 325 K at $E = 0$ to about 275 K at $E = 5$ V/nm, as shown by the inset of Fig. 4.

In conclusion, we find an unexpected electric-field-induced melting of confined monolayer ice by comprehensive molecular dynamic simulations. The critical melting temperature decreases with increasing field strength due to the field-reduced possibility of H-bond formation between the water molecules forming the monolayer ice. This melting process is attributed to the success for the electric-field-induced disordering in competition with the strong water-wall interaction induced ordering of the confined water. This novel phase transition process, which can be referenced as electromelting, is in sharp contrast to the well-known electrofreezing of water and should be important for understanding the behavior of confined water.

We are grateful to Drs. Zhuhua Zhang and Rong Shen for discussions. This work is supported by the 973 Program (2012CB933403) and the national NSF (30970557, 91023026) of China, and the Funding of Jiangsu Innovation Program for Graduate Education (CXLX11_0172) and the Fundamental Research Funds for the Central Universities.

**Reference**
[1] E.B. Moore, V. Molinero, Structural transformation in supercooled water controls the

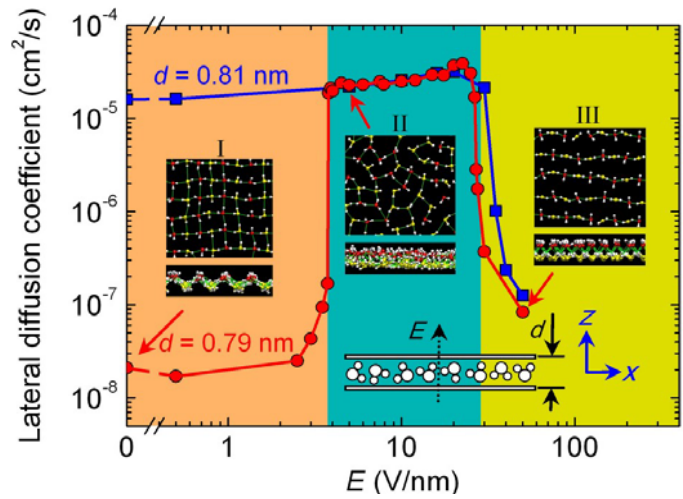

**FIG. 1** In-plane diffusion coefficient of water at 300 K as a function of the external electric field ($E$) at two interplate spacings, $d$ = 0.79 nm (red line) and 0.81 nm (blue line). Bottom inset schematically shows the simulation setup. For $d$ = 0.79 nm, the two confining plates are located at $z$ = 0.395 nm and $z$ = -0.395 nm, respectively. Top insets (I to III) show the in-plane and out-of-plane views of an ice monolayer at $E$ = 0, a liquid water phase at $E$ = 5 V/nm, and a crystal-like water phase at $E$ = 50 V/nm. White atoms represent hydrogen, while red and yellow atoms represent oxygen located above and below the midplane parallel to the confining plates, respectively. H-bonds are indicated by green lines.

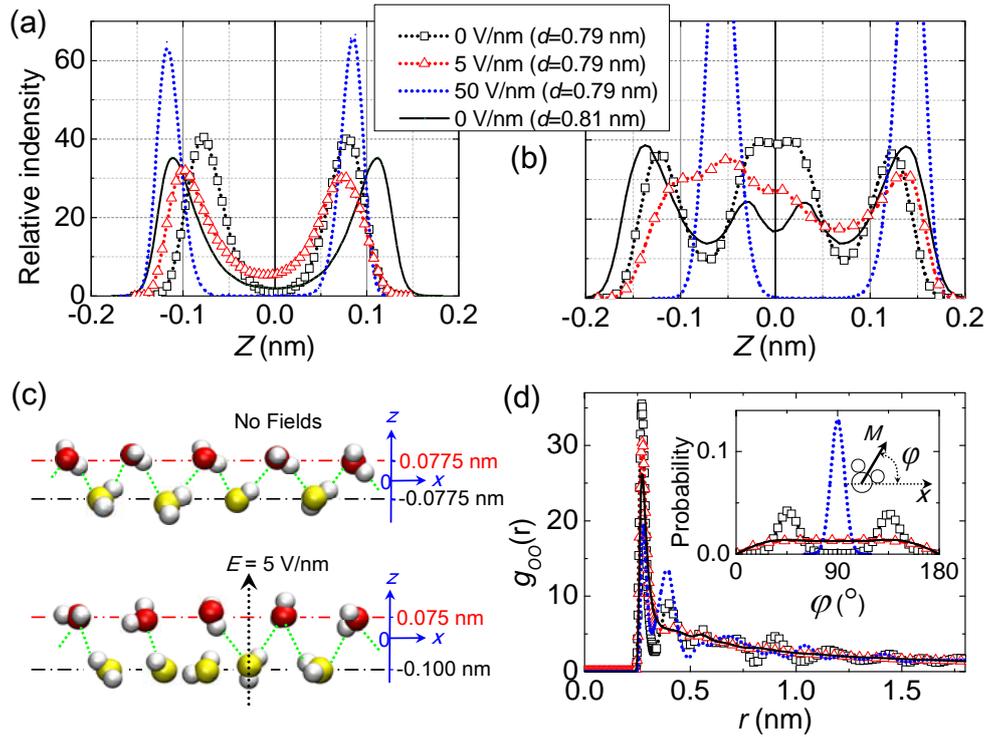

**FIG. 2** Dynamics of confined water under different fields. (a),(b) Transverse distribution of the density of (a) oxygen and (b) hydrogen atoms. (c) Schematic of the increased spacing between the two water planes caused by an electric field of 5 V/nm (bottom panel) compared to the field-free case (top panel). The color representation is the same as that given in the top inset in Fig. 1. Red and black dash-dotted lines indicate the central positions of upper oxygen plane and lower oxygen plane, respectively. (d) Oxygen-oxygen radial distribution function $g_{OO}(r)$. Inset: probability distribution for angle $\varphi$ between water molecule dipole orientation $M$ and the $x$ axis.

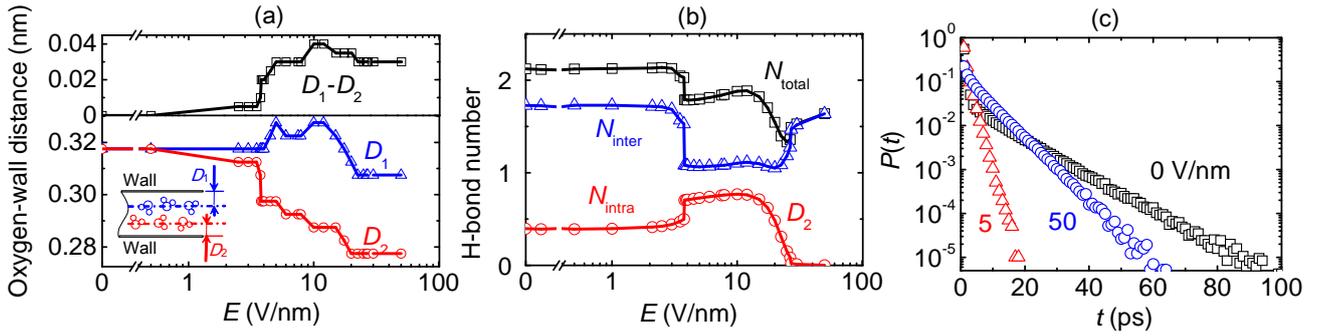

**FIG. 3** Mechanism for electromelting. (a) Calculated distance between water oxygen plane and the confining wall as a function of electric field. (b) Average number of H-bonds formed by a water molecule with other water molecules within the same plane (○), and by a water molecule with the water molecules in another plane (△) as a function of electric field, together with their sum (□). (c) Probability distribution $P(t)$ of H-bond lifetime for different field strengths.

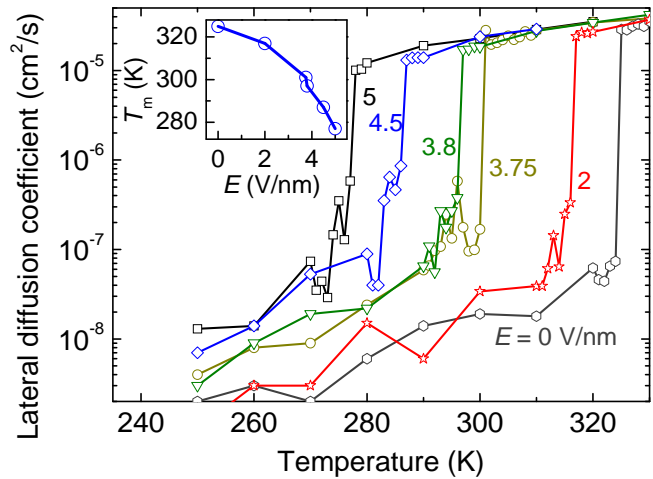

**FIG. 4** Temperature dependence of the in-plane diffusion coefficient of the confined water at $d = 0.79$ nm for different external fields $E$. The inset shows the melting point $T_m$ as a function of $E$.

Supplemental Material for

# Electromelting of Confined Monolayer Ice

Hu Qiu, Wanlin Guo

Key Laboratory for Intelligent Nano Materials and Devices of MoE and State Key Laboratory of Mechanics and Control of Mechanical Structures,
Institute of Nanoscience, Nanjing University of Aeronautics and Astronautics,
29 Yudao Street, Nanjing 210016, China

**I. Determining whether the confined water can be dissociated by perpendicular electric fields by DFT calculations.**

We performed static DFT calculations to test whether water dissociation can be induced by a perpendicular electric field in the confined water system. The initial structure of the water layer containing 64 water molecules was taken directly from our model for the MD simulations, and was then placed between two fixed graphene sheets with a separating distance of 0.77 nm. An electric field of 5 V/nm was applied perpendicular to the water plane, i.e, along +$z$ direction. After structural optimization (see below for more details of the DFT method), we found that no water dissociation occurs (Fig. S1a). Next, we intentionally dissociated a water molecule in the system into a pair of hydronium and hydroxide ions. Such dissociation raises the total energy of the system by up to 5.1 eV from initial equilibrium system. The large increase in system energy indicates that the dissociation is not favored in the confined water system under the perpendicular electric field. Actually, the dissociated ions can resume to the undissociated configuration via proton diffusing from neighboring molecules (see Fig. S1b), without any energy barrier by direct structural optimization. These results further confirm that the water dissociation is not allowed in our model under a perpendicular electric field up to 5 V/nm, and would not affect the demonstrated electromelting at a lower critical field (~3.8 V/nm).

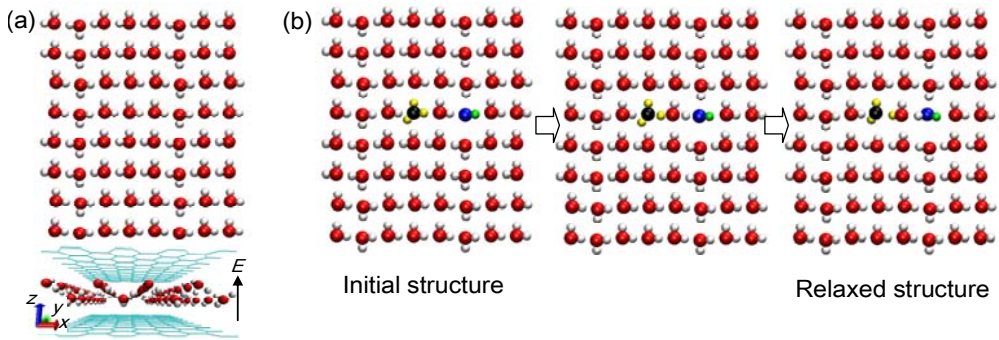

**FIG. S1** The DFT calculation model and results for confined water between two fixed graphene monolayers subjected to a perpendicular electric field of 5 V/nm. (a) The relaxed configuration of the model system in which no spontaneous water dissociation occurs. The top panel and bottom panel show the top view and side view of the unit cell, respectively, which contains a water layer and two confining graphene sheets. For clarity, graphene is not shown in the top panel. (b) Recombination process (from left to right panel) of an intentionally dissociated ion pair of hydronium (black and yellow) and hydroxide (green and blue) ions during the structural optimization of the system.

A recent ab initio MD study showed that some water molecules can be instantaneously dissociated by a field of 3.5~10 V/nm in bulk water and the dissociated ion pairs present a unique hydrogen-bonded $H_3O^+$-$H_2O$-$OH^-$ entity aligned along a direction approximatively parallel to the field [1]. This implies that water could only be dissociated when there is enough space for the $H_3O^+$-$H_2O$-$OH^-$ entity to form along the electric field. However, our confined system contains no more than two water layers along the field direction (+$z$) and thus cannot accommodate such an entity, thereby prohibiting the water dissociation.

**Methods for DFT calculations:**

All DFT calculations are performed within the framework of density-functional theory as implemented in Vienna *ab initio* simulation package (VASP) [2]. The projector augmented wave method [3] with the Perdew-Burke-Ernzerh of exchange-correlation functional [4] are employed. A kinetic-energy cutoff of 500 eV is used in the plane-wave expansion. The conjugate gradient method is used to optimize the geometry and all the atoms in the unit cell are fully relaxed until the force on each atom is less than 0.01 eV/Å. The van der Waals interactions is considered by the DFT-D2 method of Grimme [5] that adds a semi-empirical dispersion potential to the conventional Kohn-Sham DFT energy.

## II. Calculated pressure of the simulation system as a function of the electric fields.

We analyzed the pressure of our system under different electric fields and found that the pressure $P$ in our $NVT$ simulation increases gradually with increasing field in the range of 5 V/nm < $E$ ⩽ 26.5 V/nm (see Fig. S2) and then decreases to a nearly constant value for higher fields. The lateral pressure, $P_{xy}$, however, remains nearly constant for liquid water in $NP_{xy}T$ simulation at weaker fields of 3.75 V/nm < $E$ ⩽ 5V/nm and in $NVT$ simulations at stronger fields of 5 V/nm < $E$ ⩽ 26.5 V/nm. When the field is approaching 27 V/nm at which the liquid-to-solid transition occurs (see inset in Fig. S2), $P_{xy}$ jumps suddenly to ~44 bar.

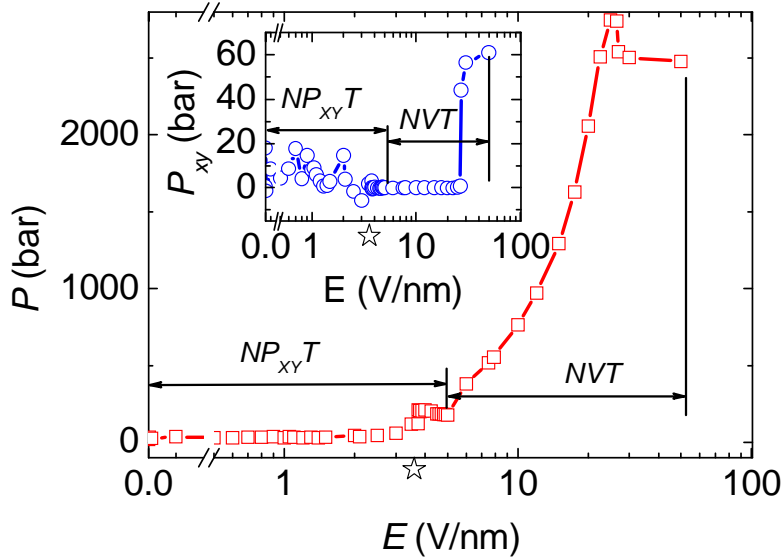

**FIG. S2** Calculated pressure $P$ and lateral pressure $P_{xy}$ (the inset) as a function of the applied electric fields. $NP_{xy}T$ ensemble was adopted for the field range of $E \leq$ 5V/nm, while $NVT$ ensemble was used for $E >$ 5 V/nm. The critical field for electromelting (~3.8 V/nm) is denoted by a star.

## III. Comparison of diffusion coefficients from TIP4P and TIP5P water models

In addition to the results using the TIP5P water model, we have conducted a series of MD simulations on the present confined system with the TIP4P water model [6]. Other simulation parameters, such as water-wall interactions, remain the same as adopted in all the rest simulations in this work. It is found that, similar to the results from the TIP5P water model, the diffusion coefficient profile for TIP4P also represents two phase transition with increasing electric field (see Fig. S3). The first transition of the system from a monolayer ice to a bilayer liquid water occurs at $E = 3.7$ V/nm, slightly weaker than that for TIP5P (3.8 V/nm). However, the critical field strength for the second phase transition (i.e., liquid-to-crystal) slightly over 40 V/nm for TIP4P is remarkably stronger than about 27 V/nm for TIP5P. Thus, the main conclusion drawn from the TIP4P water model is essentially the same as that from TIP5P.

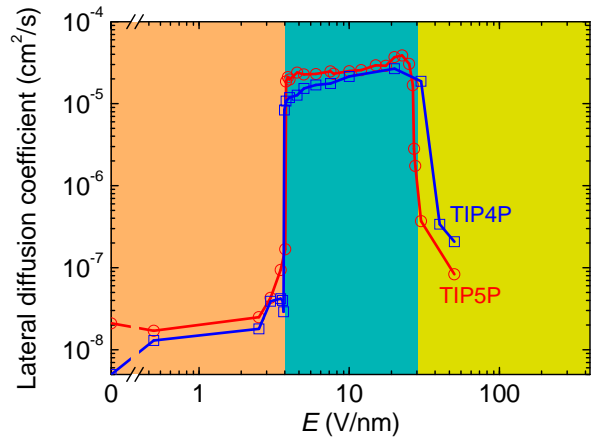

**FIG. S3** Comparison of the calculated lateral diffusion coefficients of water at 300 K as a function of the external electric field ($E$) using TIP4P and TIP5P at an interplate spacing $d = 0.79$ nm.